\documentclass[prd,twocolumn,showpacs,nofootinbibs,superscriptaddress]{revtex4}
\usepackage{epsf}
\usepackage{epsfig}
\usepackage{amsmath}

\newcommand{\nn}{\nonumber}
\newcommand{\be}{\begin{eqnarray}}
\newcommand{\ee}{\end{eqnarray}}

\renewcommand{\Im}{\text{Im}}

\begin{document}
\preprint{??}

\title{Leading Infrared Logarithms from  Unitarity, Analyticity and Crossing}

\author{Julia Koschinski}
\affiliation{Institut f\"ur
Theoretische Physik II, Ruhr--Universit\"at Bochum, D--44780
Bochum, Germany}

\author{Maxim V. Polyakov}
\affiliation{Institut f\"ur
Theoretische Physik II, Ruhr--Universit\"at Bochum, D--44780
Bochum, Germany}
\affiliation{Petersburg Nuclear Physics Institute, Gatchina, St.\
Petersburg 188300, Russia}

\author{Alexei A.  Vladimirov}
\affiliation{Institut f\"ur
Theoretische Physik II, Ruhr--Universit\"at Bochum, D--44780
Bochum, Germany}

\date{\today}

\begin{abstract}
We derive non-linear recursion equations for the leading infrared logarithms in  massless non-renormalizable
effective field theories. The derivation is based solely on the requirements of the unitarity, analyticity and crossing symmetry of the amplitudes.
That emphasizes the general nature of the corresponding equations. The derived equations allow one to compute leading infrared logarithms to essentially unlimited loop order
without performing a loop calculation. For the implementation of the recursion equation one needs to calculate tree diagrams only. The application of the equation is demonstrated
on several examples of effective field theories in four and higher space-time dimensions.
\end{abstract}


 \maketitle

\section{Introduction}

Effective Field Theories (EFTs) are {\it non-renormalizable} field
theories, which allow the investigation of the infrared (low-energy) behaviour of
various physical systems (see recent review \cite{Weinberg}).
The standard tool for the studies of the asymptotic behaviour of {\it renormalizable}
field theories is the method of renormalization group equations (RGEs).
In the case of EFTs the method of RGEs must be modified  as the number of counterterms
increases rapidly with the loop order.

A possibility of the systematic construction of RGEs for non-renormalizable quantum field theories
was demonstrated  in Ref.~\cite{BC_2003}. In particular, it was shown that the
series of the leading logarithms (LLs) can be obtained by calculation of one loop diagrams.
However, the solution of the RGEs derived in Ref.~\cite{BC_2003}  requires the calculation
of non-trivial one-loop diagrams, the number of which is rapidly increases with the loop order. Therefore, the implementation of this method in practice is not
an easy task.
The method of Ref.~\cite{BC_2003} has been applied  in Ref.~\cite{bijnens} for the calculation of the five-loop LLs for the pion mass
in the massive $O(N+1)/O(N)$ sigma model.
In Ref.~\cite{Biss2} the authors, using dispersive methods, calculated the three-loop LLs
to $\pi\pi$ scattering in massless Chiral Perturbation Theory (ChPT).

Recently, a completely different method for the
calculation of LLs in a wide class
of non-renormalizable {\it massless} field theories was developed in Refs.~\cite{MKV_LLog,MKV_FF}.
The { non-linear recursion} equations derived in Refs.~\cite{MKV_LLog,MKV_FF} allow one to obtain the LLs contributions
without performing non-trivial loop calculations at each loop order.

In the present paper we show that the non-linear recursion equation for
LLs is  a consequence of analyticity, unitarity, and crossing symmetry of the
$S$-matrix. The requirement of
analyticity, unitarity, and crossing symmetry for massless particles
allows to obtain LLs coefficients without calculation of loop integrals.
[We note that  these requirements in conformal field theories
lead to very powerful bootstrap equations, see \cite{Polyakov}.]
The form of the corresponding equations is such that one can easily generalize them in
various ways -- for arbitrary space-time dimension, different
symmetry groups, different spins of the fields etc.

To make the presentation of our method transparent we shall restrict ourselves to a massless scalar EFT, which
is described by the following generic action:
\begin{eqnarray}\label{L}
S=\int d^Dx\ \left[ \frac{1}{2}\partial_\mu \phi^a \partial^\mu\phi^a -V(\phi,\partial\phi)\right],
\end{eqnarray}
where the expansion of the interaction part of the action $V(\phi,\partial\phi)$ starts with four fields $\phi$,
the corresponding ``$\phi^4$" part contains $2 k$ derivatives.
The index $a$ (it can be a multi-index as well)  corresponds to a possible internal symmetry of  theory (\ref{L}).
EFTs of type (\ref{L}) have
the property that the anomalous dimension of fields is zero
at the leading order. The
theories with ``$\phi^3$" interaction can be considered in
similar framework, this a more complicated case will be
considered elsewhere. We stress that the absence of mass
is crucial for our discussion below.

\section{General method}
\label{sec:general}

We consider the 4-particle
scattering amplitude $\phi^a+\phi^b\to \phi^c+\phi^d$ in four dimensional space (generalization to an arbitrary number of dimensions is given in Section~\ref{arbD}).
The
LLs for other physical quantities, such as form factors \cite{MKV_FF}, (generalized) parton distributions \cite{MKV_GPD}, etc., are
related to the LLs of the 4-point amplitude.

The  $2\rightarrow2$ amplitude can be decomposed in the irreducible representations
of an internal symmetry group of the EFT (\ref{L}):
\begin{eqnarray}\label{<pp|pp>}
A^{abcd}(s,t,u)=\sum_I P_I^{abcd}A^I(s,t,u).
\end{eqnarray}
Here $P_I^{abcd}$ is a projector on an invariant subspace corresponding to the
irreducible representation $I$ of an internal symmetry group.
According to the Wigner-Eckart
theorem  the corresponding projectors can be obtained as  the convolution of two Clebsch-Gordon
coefficients, see e.g.\cite{Cvitanovich}.

The renormalizability of a theory depends on the
dimension of the coupling constants that enter  action (\ref{L}).
Let us consider the case of
the action (\ref{L}) with one coupling constant of dimension
$(4-D-2 k)$ ($2 k$ is the number of derivatives in the interaction part of the Lagrangian, $D$ is the dimension of space-time).
We denote the corresponding coupling constant as $1/F^2$, the dimension of the constant $F$ is $k+D/2-2$ (the constant $F$ corresponds to the pion
decay constant in usual ChPT). The case of several couplings
we consider in Section~\ref{secsev}. At $k=2-D/2$ theory (\ref{L}) is renormalizable, and at $k>2-D/2$ it
is a non-renormalizable low-energy EFT.

We shall work with the partial wave amplitudes (see all definitions in Appendix~A) as they depend only
on one energy variable $s$. For $D=4$ (see generalization for an arbitrary space-time dimension in Section~\ref{arbD})
the partial wave amplitude in LL approximation  can be represented in the most general form as follows:
\begin{widetext}
\begin{eqnarray}\label{t^I_l}
 t^I_l(s)=\frac{\pi}{2}\sum_{n=1}^\infty \frac{\hat S^n}{2l+1}
\sum_{i=0}^{n-1}\alpha^{I,l}_{n,i}\ln^{i}\Big(\frac{\mu^2}{s}\Big)\ln^{n-i-1}\Big(\frac{\mu^2}{-s}\Big)
+\mathcal{O}(\text{NLL}),
\end{eqnarray}
\end{widetext}
where $\hat S=\frac{s^k}{(4\pi F)^2}$ is a dimensionless expansion parameter, $F^2$ is a
coupling in the Lagrangian. The scale parameter $\mu$ is arbitrary in the LL approximation, since its change influences
the next-to-leading logs only.
$\mathcal{O}(\text{NLL})$  stands for terms with
next-to-leading logarithms (NLL). The series of LLs for the amplitude (\ref{t^I_l}) takes into account that the scattering amplitude
has both right and left cuts in the complex plane of the Mandelstam variable $s$.

Our aim is to derive the recursion relations for the coefficients in
front of LLs in the low-energy expansion of the partial wave amplitudes:
\begin{eqnarray}\label{t^I_lomega}
 t^I_l(s)=\frac{\pi}{2}\sum_{n=1}^\infty \omega^{I}_{nl}\ \frac{\hat S^n}{2l+1}
\ln^{n-1}\Big(\frac{\mu^2}{s}\Big)
+\mathcal{O}(\text{NLL}).
\end{eqnarray}
Obviously one has:
\begin{eqnarray}\label{sum_a=w}
\omega_{nl}^I=\sum_{i=0}^{n-1}\alpha^{I,l}_{n,i},
\end{eqnarray}
where the index $n$ corresponds to the loop order plus one, and the angular momentum $l$ is restricted to $l\leq k n$.
$I$ denotes the irreducible representation
of the corresponding internal symmetry group.

In order to derive the recursive equations for the LL coefficients $\omega_{nl}^I$ we shall employ the
unitarity, analyticity and crossing symmetry of the scattering amplitude.

The crossing symmetry relates the amplitudes with the interchanged momenta to the amplitudes of different symmetry group representations (the summation over
the repeated ``representation indices" is always assumed):
\begin{eqnarray}\label{Crossing}
A^I(s,t,u)&=&C^{II'}_{tu}A^{I'}(s,u,t),\\
\nn
A^I(s,t,u)&=&C^{II'}_{su}A^{I'}(u,t,s).
\end{eqnarray}
The crossing matrices $C$ are defined as
\begin{eqnarray}\label{def_C}
C^{II'}_{tu}=P_{I}^{abcd}P_{I'}^{bacd}\frac{1}{d_I},~~C^{II'}_{su}=P_{I}^{abcd}P_{I'}^{bdac}\frac{1}{d_I},
\end{eqnarray}
where $d_I$ is the dimension of the irreducible representation $I$ (see the explicit expressions for the crossing matrices below).

The partial wave scattering amplitudes possess right and left cuts in the complex plane of the Mandelstam variable $s$. The
right cut discontinuity is fixed by the unitary relation:
\begin{eqnarray}\label{Unitarity}
\Im\, t_l^I(s)=|t^{I}_l(s)|^2 +\mathcal{O}(\text{Inelastic part}).
\end{eqnarray}
The unitarity relation is valid for the physical region $s>0$.
Under the inelastic part we understand the part with more than two
particles in an intermediate state. Obviously the LLs coefficients
do not depend on this part of the expression.
Indeed, by cutting  more than two lines in any diagram the power of logarithms decreases and hence such cuts influence only NLLs.
We can restrict ourselves to the elastic unitarity relation, since
only LL coefficients are considered.

Substituting the LL expression (\ref{t^I_l}) for $t_l^I(s)$ into
the unitarity relation (\ref{Unitarity})
and collecting the coefficients in front of
$\hat S^n L^{n-2}$ ($L\equiv \ln\left(\mu^2/s\right)$), we find the relation
\begin{eqnarray}\label{from_Unitarity}
\sum_{i=0}^{n-i-1}(n-i-1)\alpha^{I,l}_{n,i}=\frac{1}{2(2l+1)}\sum_{i=1}^{n-1}\omega_{il}^{I}\omega_{n-i,l}^{I}.
\end{eqnarray}
Thus, the unitarity allows us to relate  $\alpha$ and $\omega$ coefficients. However, corresponding relations (\ref{from_Unitarity}) do not
allow us to obtain the closed form equation for the coefficients $\omega$.

We need some additional information about the left cut. This
information is provided by the dispersion relations (analyticity) and the crossing
symmetry (\ref{Crossing}). One can show that if only two-particle cuts  are taken into account, the following relation connects
discontinuities on the left and right cuts (see
derivation in the Appendix~A)
\begin{widetext}
\begin{eqnarray}\label{ImRoy}
\Im \,t^I_l(s)=\sum_{l'=0}^\infty C_{su}^{II'}\frac{2(2l'+1)}{s}\int_0^{-s}
ds' \,
P_l\Big(\frac{s+2s'}{-s}\Big)P_{l'}\Big(\frac{2s+s'}{-s'}\Big)\Im\,
t_{l'}^{I'}(s').
\end{eqnarray}
\end{widetext}
This relation for the $\pi\pi$ scattering is well-known, it is a
consequence of the Roy equation \cite{Roy}, it was used in Ref.~\cite{Biss2} to calculate
three-loop LLs in ChPT.
Substituting Eq.~(\ref{t^I_l}) into Eq.~(\ref{ImRoy})
and collecting the terms in front of the LLs  we
obtain  the following relation in addition to (\ref{from_Unitarity}):
\begin{eqnarray}\label{from_ImRoy}
\sum_{i=0}^{n-i-1}i\alpha^{I,l}_{n,i}=\sum_{l'=0}^{k n} \frac{C^{II'}_{su}}{2l'+1}\sum_{i=1}^{n-1}\omega_{il'}^{I'}\omega_{n-i,l'}^{I'}
(-1)^{l+l'}\Omega_{kn}^{l'l},
\end{eqnarray}
where the matrices $\Omega^{ll'}_{n}$ are defined as follows
\begin{eqnarray}
\nn
\Big(\frac{z-1}{2}\Big)^nP_{l}\Big(\frac{z+3}{z-1}\Big)=\sum_{l'=0}^n
\Omega_{n}^{ll'}P_{l'}(z).
\end{eqnarray}
The $(n+1)\times(n+1)$ matrices $\Omega^{ll'}_{n}$ perform the crossing transformation for the partial waves.
Main properties and explicit expressions for the matrices $\Omega^{ll'}_{n}$
are given in the Appendix~B.

Summing relations (\ref{from_Unitarity}) and (\ref{from_ImRoy})
and taking into account $t\leftrightarrow u$ crossing symmetry we arrive
at the following closed non-linear recursive relation for LL coefficients $\omega_{nl}^I$:
\begin{widetext}
\begin{eqnarray}\label{w=ww}
\omega^I_{nl}=\frac{1}{n-1}\sum_J \sum_{i=1}^{n-1}\sum_{l'=0}^{kn}\frac 12\ \Big(\delta^{ll'}\delta^{IJ}+
C_{st}^{IJ} \Omega_{kn}^{l'l}
+C_{su}^{IJ}
(-1)^{l+l'}\Omega_{kn}^{l'l}\Big)\frac{\omega^{J}_{il'}\omega^{J}_{n-i,l'}}{2l'+1}.
\end{eqnarray}
\end{widetext}
[Here we introduce the matrix $C_{st}$ which is defined as a product of the crossing matrices (\ref{def_C}) $C_{st}=C_{su} C_{tu}C_{su}$].
The recursion relation (\ref{w=ww}) for the LL coefficients should be supplemented by initial conditions, {\it i.e.}
by the values of $\omega_{nl}^I$ at $n=1$ and correspondingly $l=0,1$. The corresponding values can be obtained by the {\it tree level}
calculation of the partial wave amplitudes $t_l^I(s)$ (see Eq.~(\ref{t^I_lomega})) using  action (\ref{L})
of an EFT under consideration.

The recursion equation (\ref{w=ww}) is the main result of the present paper. This equation allows the calculation
of the leading infrared logarithms to an essentially unlimited order.
Furthermore, this method presents a puissant tool
for the study of the general structure of the infrared logarithms.
For the derivation of Eq.~(\ref{w=ww}) we used only the unitarity, analyticity, and crossing symmetry of the amplitude.
This fact emphasizes a general nature of the non-linear recursion equation (\ref{w=ww}).
The corresponding equation
can be derived for many physical problems described by a
non-renormalizable effective low-energy Lagrangian, e.g.
theory of critical phenomena, low-energy quantum gravity,
theory of magnetics, etc.

In the next Sections we apply our method to several EFTs. Also we give the generalization of the method on
an arbitrary dimension and on the theories with mixed renormalizable
and non-renormalizable interactions.\\

\section{4D massless $O(N+1)/O(N)$ $\sigma$-model.}
\label{sec:sigma}

We have several reasons to consider the $O(N+1)/O(N)$ $\sigma$-model.
First, this model was considered in \cite{MKV_LLog} where the
recursion relations for LLs were derived by a completely different method.
Second, this model (at $N=3$) is equivalent to the Weinberg
Lagrangian \cite{weinberg} of ChPT.
The Lagrangian of the $O(N+1)/O(N)$ $\sigma$-model has the form
\begin{widetext}
\begin{eqnarray}\label{ChPT}
\mathcal{L}=\frac{1}{2}\big(\partial_\mu\sigma\partial^\mu\sigma+\partial_\mu
\phi^a
\partial^\mu
\phi^a\big)=\frac{1}{2}\partial_\mu\phi^a\partial^\mu\phi^a-\frac{1}{8F^2}(\phi^a\phi^a)\partial^2(\phi^b\phi^b)+\mathcal{O}(\phi^6),
\end{eqnarray}
\end{widetext}
where $\sigma^2=F^2-\sum_{a=1}^N\phi^a\phi^a$. The $\mathcal{O}(\phi^6)$ part does not
contribute to the 4-particle amplitude due to the absence of
masses. The $\sigma$-model (\ref{ChPT}) belongs to the here considered class of EFTs (\ref{L}) with $D=4$ and $k=1$.

The theory (\ref{ChPT}) possesses $O(N)$ symmetry, therefore the expansion in invariant tensors (\ref{<pp|pp>})
goes over the irreducible representations of the $O(N)$ group. Namely, over the three representations of dimensions
$d_I=\left\{1,\frac{N(N-1)}{2},\frac{(N+2)(N-1)}{2}\right\}$. We enumerate these representations by index $I=0,1,2$;
such naming corresponds to the value of the isospin in the s-channel for the case $N=3$.

The corresponding projection operators on these representations have the following form:
\begin{widetext}
\begin{eqnarray}\nn
P^{abcd}_0=\frac{1}{N}\delta^{ab}\delta^{cd}~, ~~
P^{abcd}_1=\frac{1}{2}\Big(\delta^{ac}\delta^{bd}-\delta^{ad}\delta^{bc}\Big)~,
~~
P^{abcd}_2=\frac{1}{2}\Big(\delta^{ac}\delta^{bd}+\delta^{ad}\delta^{bc}\Big)-\frac{1}{N}\delta^{ab}\delta^{cd}.
\end{eqnarray}
\end{widetext}
The straightforward calculation with help of Eq.~(\ref{def_C}) gives the crossing matrices
\begin{eqnarray}\nn
C_{tu}=\begin{pmatrix}
1 & 0 & 0 \\
0 & -1& 0 \\
0 & 0 & 1 \end{pmatrix},~~ C_{su}=\begin{pmatrix}
  \frac{1}{N} & -\frac{N-1}{2} & \frac{N^2+N-2}{2N} \\
  -\frac{1}{N} & \frac{1}{2} & \frac{N+2}{2N} \\
  \frac{1}{N} & \frac{1}{2} & \frac{N-2}{2N}
\end{pmatrix}.
\end{eqnarray}
Substituting the crossing matrices into Eq.~(\ref{w=ww}) we obtain explicit form of recursion relations for the LL coefficients. The initial conditions for
the recursion can be obtained by the trivial tree-level calculation of the scattering amplitude with the help of the Lagrangian (\ref{ChPT}).
The result is $\omega_{10}^{I=0}=N-1$, $\omega_{10}^{I=2}=-1$, and $\omega_{11}^{I=1}=1$, all other coefficients $\omega_{1l}^I$ are zero.

At first glance, the recursion relation (\ref{w=ww}) is different from that derived in Ref.~\cite{MKV_LLog}. The
reason is that the LL coefficients $\omega_{nl}^I$ with fixed $O(N)$ quantum numbers are not independent and
can be expressed in terms of the universal LL coefficients $\omega_{nl}$ which were used in Ref.~\cite{MKV_LLog}.
The corresponding relations have the following form:

\begin{widetext}
\begin{eqnarray}\label{w=wI}
\omega^{I=0}_{nl}&=&\sum_{l'=0}^n \omega_{nl'}\Bigg[N
\delta^{ll'}+\Big((-1)^{l'}+(-1)^{l}\Big)\Omega_n^{l'l}\Bigg],~~
\\ \nn
\omega^{I=1}_{nl}&=&\sum_{l'=0}^n
\omega_{nl'}\Big((-1)^{l'}-(-1)^{l}\Big)\Omega_n^{l'l},~~
\omega^{I=2}_{nl}=\sum_{l'=0}^n
\omega_{nl'}\Big((-1)^{l'}+(-1)^{l}\Big)\Omega_n^{l'l}.
\end{eqnarray}
\end{widetext}
The origin of such relations is very simple -- the $O(N+1)/O(N)$ $\sigma$ model is a theory
with one coupling constant and hence the channels with different internal symmetry quantum numbers
are interrelated.

\begin{figure}
\includegraphics[width =8.cm]{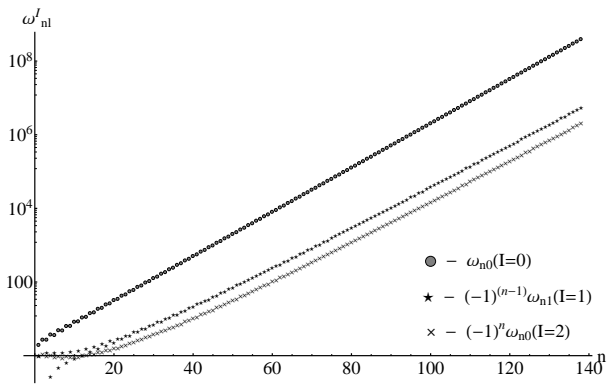}
\caption{The lowest partial wave LL coefficients in the $O(4)/O(3)$ sigma model up to the 137th loop order.}
\label{fig:halla}
\end{figure}

The equation on $\omega$ is obtained  from (\ref{w=ww})
by the substitution (\ref{w=wI}), it reads
\begin{widetext}
\begin{eqnarray}\label{O(N)_w}
\omega_{nj}=\frac{1}{n-1}\sum_{m=1}^{n-1}\sum_{i=0\atop \scriptstyle{\rm even}}^{m}\sum_{l=0\atop \scriptstyle{\rm even}}^{n-m}
B_{j}^{(m,i)(n-m,l)} \omega_{mi}\omega_{(n-m) l},
\end{eqnarray}
\end{widetext}
with the initial condition $\omega_{10}=1, \omega_{11}=0$ .The coefficients   $B_{j}^{(m,i)(n-m,l)}$ are given by:
\begin{widetext}
\be
\label{O(n)beta}
B_{j}^{(m,i)(p,l)}=\frac{1}{2j+1}\left[\frac{N}{2}\delta_{ij}\delta_{lj}+\delta_{ij} \Omega^{li}_{p}
+\delta_{l j} \Omega^{il}_{m}
\right]
+\left(1+(-1)^j\right)\sum_{k=0}^{{\rm min }[p,m]} \frac{\Omega_m^{ik} \Omega_{p}^{lk} \Omega_{m+p}^{kj}}{2 k+1}\, .
\ee
\end{widetext}
Eq.~(\ref{O(N)_w}) coincides exactly with the equation obtained  by a completely different, more complicated method in Ref.~\cite{MKV_LLog}.
In the present paper the equation was obtained
from a simple symmetry consideration. The detailed discussion of the properties of the recursion equation (\ref{O(N)_w}) (and hence of Eq.~(\ref{w=ww}))
and its particular solutions can be found in Ref.~\cite{MKV_LLog}. For convenience of the reader we present in Appendix~C [Tables~I-III]
the values of the LL coefficients $\omega_{nl}^I$ up to the 4-loop order. In Fig.~\ref{fig:halla} we show the values for the lowest partial waves LL coefficients
in the $O(4)/O(3)$ sigma model up to the 137th loop order.

\section{4D $O(N)$ symmetric $\phi^4$ theory}
\label{sec:fi^4}

As a next example we consider a renormalizable field theory. For
simplicity we consider the same $O(N)$ symmetry as in the previous case. A
possible renormalizable Lagrangian of type (\ref{L}) is:
\begin{eqnarray}\label{phi^4}
\mathcal{L}=\frac{1}{2}\partial_\mu\phi^a\partial^\mu\phi^a-\frac{1}{8 F^2}(\phi^a\phi^a)^2.
\end{eqnarray}
Due to the same $O(N)$ symmetry as for the $\sigma$-model (\ref{ChPT}) the crossing matrices
are the same. The differences to the previous case are the number of derivatives in the interaction vertex,
$i.e.$  $k=0$ in Eqs.~(\ref{t^I_lomega},\ref{w=ww}) and the coupling constant $F$ is dimensionless now.

The low-energy expansion of the scattering amplitude in the renormalizable theory (\ref{phi^4}) goes
over increasing power of LLs (note that $\hat S=1/(4\pi F)^2$ at $k=0$ in Eq.~(\ref{t^I_lomega})):

\begin{eqnarray}\label{t^Ifi4}
 t^I_l(s)=\frac{\pi}{2}\sum_{n=1}^\infty \frac{\omega^{I}_{nl}}{2l+1} \frac{1}{(4\pi F)^{2n}}
\ln^{n-1}\Big(\frac{\mu^2}{s}\Big).
\end{eqnarray}
The LL coefficients $\omega_{nl}^I$ satisfy the recursion relation (\ref{w=ww}) with $k=0$.
The initial conditions are obtained by calculation of the scattering amplitude at the tree level in
theory (\ref{phi^4})-- $\omega_{10}^{I=0}=-(N+2)$, $\omega_{10}^{I=2}=-2$ with all other $\omega_{1l}^I$ are zero.

Eq.~(\ref{w=ww})  is simplified considerably at $k=0$ as it contains only the matrix $\Omega_0^{l´l}=\delta^{l´0}\delta^{l0}$. This implies that
the LL coefficients are restricted only to the lowest partial wave $l=0$. Using this simplification and introducing the following notation
$\omega_n^I\equiv \omega_{n0}^I$ we can rewrite the recursion relation (\ref{w=ww}) for the renormalizable theory (\ref{phi^4}) in a simpler form:
\be
\omega_n^{I=0}&=&\frac{1}{n-1}\sum_{i=1}^n \left[ \frac{N+2}{2 N}\ \omega_{i}^{I=0} \omega_{n-i}^{I=0} +\right. \\
\nn
&+& \left. \frac{N^2+N-2}{2N}\ \omega_{i}^{I=2} \omega_{n-i}^{I=2}\right] ,\\
\omega_n^{I=1}&=&\frac{1}{n-1}\sum_{i=1}^n  \omega_{i}^{I=1} \omega_{n-i}^{I=1},\\
\omega_n^{I=2}&=&\frac{1}{n-1}\sum_{i=1}^n \left[ \frac{1}{N}\ \omega_{i}^{I=0} \omega_{n-i}^{I=0} +\right. \\
\nn
&+& \left. \frac{N-1}{N}\ \omega_{i}^{I=2} \omega_{n-i}^{I=2}\right].
\ee
This system of equations can be easily solved with the following result:
\be
\label{solution}
\omega_{n}^{I=0}&=&\frac{N+2}{2}\ \omega_{n}^{I=2}=-(N+2) \left(-\frac{N+8}{2}\right)^{n-1}, \\
\nn
\omega_{n}^{I=1}&=&0.
\ee
Substituting this solution into the LL expansion of the scattering amplitudes (\ref{t^Ifi4}), we can
perform the summation of the LLs with the result:
\be
\label{RGphi4}
t_0^{I=0}(s)&=&-\frac{\pi}{2}\ \frac{N+2}{(4\pi F)^2}\ \frac{1}{1+\frac{b_1}{(4\pi F)^2}\ln\left(\frac{\mu^2}{s}\right)}, \\
\nn
t_0^{I=2}(s)&=&-\frac{\pi}{2}\ \frac{2}{(4\pi F)^2}\ \frac{1}{1+\frac{b_1}{(4\pi F)^2}\ln\left(\frac{\mu^2}{s}\right)}.
\ee
All other amplitudes are zero in the LL approximation. In Eq.~(\ref{RGphi4}) the constant $b_1=\frac{N+8}{2}$ is the well known result
for the one-loop beta function of the $O(N)$ symmetric $\phi^4$ theory (\ref{phi^4}), see e.g.
Ref.~\cite{KVT_v^4}.

The logs enter  Eq.~(\ref{RGphi4}) in the combination in which one can easily recognize the
running  coupling constant (denoted as $1/F^2$ in this paper)
for the $O(N)$ symmetric $\phi^4$ theory. It is not surprising. Indeed,
for a generic renormalizable theory (e.g. $k=0, D=4$) the recursion equation (\ref{w=ww}) is reduced to the following form:
\be
\label{w=wwren}
\omega_n^I= \frac{1}{n-1}\sum_{i=1}^{n-1}\sum_J B^{IJ}\ \omega_{n-i}^J\omega_i^J,
\ee
where the matrix $B^{IJ}$ is independent of the index $n$ and it is expressed in terms of the crossing matrices:
\be
B^{IJ}=\frac 12 \left(1+C_{st}+C_{su}\right)^{IJ}.
\ee
The recursion equation (\ref{w=wwren}) can be reduced to a simple system of differential equations. For that we introduce a generating function
for the LL coefficients $\omega^I_n$ as follows $\omega^I(t)\equiv \sum_{n=1}^{\infty} \omega_n^I\ t^{n-1}$. Then the recursion relation (\ref{w=wwren}) is reduced
to the following differential equation:

\be
\label{diffRGE}
\frac{d \omega^I(t)}{dt}=\sum_J B^{IJ}\ \omega^J(t) \omega^J(t),
\ee
with the initial conditions that can be obtained by the calculation of the tree diagrams, e.g. in $O(N)$ symmetric $\phi^4$ theory $\omega^I(0)=\left\{-N-2,0,-2\right\}$.
Eq.~(\ref{diffRGE}) has the form typical for the RGEs for the running coupling constant. In this way, we demonstrate that our recursion relation (\ref{w=ww})
for LL coefficients is reduced to the standard RGEs in the case of a renormalizable field theory. Eq.~(\ref{w=ww}) can be considered also as the generalization of
RGEs for the case of non-renormalizable EFTs.

\section{Theory with renormalizable and non-renormalizable
interactions}
\label{secsev}

In this section we consider
the theory with mixed renormalizable
and non-renormalizable interactions.  To be specific we consider a Lagrangian
which is the sum of the $O(N+1)/O(N)$ $\sigma$-model and the $O(N)$ symmetric
$\phi^4$ theory:

\begin{widetext}
\begin{eqnarray}\label{ChPT+sb}
\mathcal{L}_{2}=\frac{1}{2}\big(\partial_\mu\sigma\partial^\mu\sigma+\partial_\mu
\phi^a
\partial^\mu
\phi^a\big)-\frac{\lambda}{8}\ (\phi^a\phi^a)^2=\frac{1}{2}\partial_\mu\phi^a\partial^\mu\phi^a-\frac{1}{8F^2}(\phi^a\phi^a)\partial^2(\phi^b\phi^b)
-\frac{\lambda}{8}\ (\phi^a\phi^a)^2+\mathcal{O}(\phi^6),
\end{eqnarray}
\end{widetext}
where $\sigma^2=F^2-\sum_{a=1}^N\phi^a\phi^a$.
In this case the expansion is performed over two parameters: a dimensionless
coupling $\lambda$ and dimensional $1/F^2$. We note that at $N=3$ the theory (\ref{ChPT+sb}) corresponds to the famous two-flavour
chirally symmetric Weinberg Lagrangian plus the chiral symmetry breaking term proportional
to the constant $\lambda$. That chiral symmetry breaking term is proportional to the constant $l_3$ in the notations of Gasser and Leutwyler \cite{Gasser}. However,
here we force this term to be of the same order as the Weinberg Lagrangian.

The  series in infrared  LLs
for the amplitude in the theory (\ref{ChPT+sb})  can be presented in the form of double expansion in the parameters $1/F^2$ and $\lambda$:

\begin{eqnarray}\label{t^I_lomegaRNR}
 t^I_l(s)=\frac{\pi}{2}\sum_{n,m=0}^\infty \omega^{I}_{nml}\ \frac{\hat S^n}{2l+1}\ {\hat \lambda}^m
\ln^{n+m-1}\Big(\frac{\mu^2}{s}\Big).
\end{eqnarray}
Here $\hat S=\frac{s}{(4\pi F)^2}$ and $\hat \lambda=\lambda/(4\pi)^2$ are dimensionless expansion parameters.
The LL coefficients $\omega^{I}_{nml}$ now have an additional index $m$, which shows the order in the constant $\lambda$.
Also, for convenience, we define $\omega^{I}_{00l}\equiv 0$. Obviously, $\omega_{n0l}^I$ coincide with the LL coefficients
for the $O(N+1)/O(N)$ $\sigma$-model considered in Section~\ref{sec:sigma}, and correspondingly the coefficients $\omega_{0ml}^I$ are equal
to the LL coefficients for the $O(N)$ symmetric $\phi^4$ theory (see Section~\ref{sec:fi^4}).

Imposing the conditions of the unitarity, analyticity and crossing symmetry on the LL amplitude (\ref{t^I_lomegaRNR})
and repeating steps described in Section~\ref{sec:general}, we arrive at the following recursion equation for LL coefficients
$\omega_{nml}^I$:

\begin{widetext}
\begin{eqnarray}\label{w=ww(R+NR)}
\omega^I_{nml}=\frac{1}{(n+m-1)}\sum_J \sum_{i,j=0}^{n,m}\sum_{l'=0}^{n}\frac 12 \Big(\delta^{ll'}\delta^{IJ}+
C_{st}^{IJ} \Omega_{n}^{l'l}
+C_{su}^{IJ}
(-1)^{l+l'}\Omega_{n}^{l'l}\Big)\frac{\omega^{J}_{ijl'}\omega^{J}_{n-i,m-j,l'}}{2l'+1}.
\end{eqnarray}
\end{widetext}
As usual, the initial conditions for this recursion equation can be obtained by the tree level calculation
of the scattering amplitude with the Lagrangian (\ref{ChPT+sb}). Actually, the initial conditions are the combination
of the corresponding initial conditions discussed in the previous two sections.

An important feature of the recursion equation (\ref{w=ww(R+NR)}) is that its dependence on the index $m$ is trivial. Eq.~(\ref{w=ww(R+NR)}) with fixed
indices $n$ and $l$ has the generic form (\ref{w=wwren}).
Therefore, the corresponding
equation, in principle, can be solved for fixed indices $n$ and $l$.
Introducing the generating function:
\be
\omega_{nl}^I(t)=\sum_{m=0}^\infty \omega_{nml}^{I}\ t^m\ ,
\ee
we can partially sum up the LL expansion of the amplitude (\ref{w=ww(R+NR)}):
\be
t^I_l(s)=\frac{\pi}{2}\sum_{n=0}^\infty \omega^{I}_{nl}\left(\hat \lambda\ \ln\Big(\frac{\mu^2}{s}\Big)\right)\ \frac{\hat S^n}{2l+1}\
\ln^{n-1}\Big(\frac{\mu^2}{s}\Big).
\ee
This expansion is similar to the LL series for the theory with a single coupling constant (\ref{t^I_lomega}), the only difference is that
the LL coefficients $\omega_{nl}^I$ are functions of  $\lambda$ and the log.
Expressions for the functions $\omega_{nl}^I(t)$ are rather complicated, therefore we give only a couple of (non-trivial) examples for
$N=3$, corresponding to ChPT. We consider the case of $n=2$, that corresponds to the one-loop chiral log which is dressed by any number of loops
with a $\lambda\phi^4$ interaction. The result of simple calculations gives:
\be
\nn
\omega_{20}^{I=0}(t)&=&\frac{950}{663} f_1(t)-\frac{160}{117} f_2(t)-\frac{10}{153} f_3(t),\\
\label{chlm}
\omega_{20}^{I=2}(t)&=&\frac{140}{663} f_1(t)-\frac{40}{117} f_2(t)+\frac{20}{153} f_3(t),\\
\nn
\omega_{21}^{I=1}(t)&=&-\frac{100}{221} f_1(t)+\frac{10}{39} f_2(t)+\frac{10}{51} f_3(t),
\ee
with the functions $f_i(t)$ given by:
\be
f_i(t)=\frac{1}{t (1+b_1 t)^{\gamma_i}},
\ee
where $b_1=11/2$ is the coefficient of the one-loop $\beta$-function of the $O(3)$ symmetric $\phi^4$ theory.
The ``anomalous dimensions" $\gamma_i$ have the following values: $\gamma_1=9/11,\gamma_2=40/33,\gamma_3=10/33$.
Using the result (\ref{chlm}) one can see that the inclusion of the $\phi^4$ interaction to the massless chiral
Lagrangian ($O(4)/O(3)$ $\sigma$-model) leads to decreasing of the coefficients in front of the one-loop chiral log (except
the $I=1$ amplitude for which the one loop chiral log is zero). Possible physics implication of this will be considered elsewhere.

\section{Effective field theories in arbitrary dimension}
\label{arbD}
The derivation of the recursion equation (\ref{w=ww}) was done for $D=4$.
It is straightforward to generalize it for an arbitrary even
dimension $D>4$ \footnote{The case of $D=2$ requires a special consideration, since in this case the partial wave expansion
degenerates to forward and backward scattering only. This interesting case will be considered elsewhere.}. We restrict ourselves to the even dimensions in order to avoid
the appearance of power-like singularities
in the amplitudes. The analytic structure of the amplitude
in an odd dimensional theory should be investigated separately.

We consider the Lagrangian of type (\ref{L}) with the interaction part
containing $2k$ derivatives. For $D>4$ all theories of type (\ref{L})
are non-renormalizable. Simple dimensional analysis shows that
an $(n-1)$-loop diagram in a theory (\ref{L}) scales with low external momenta
as $\sim (p^2)^{kn+(n-1)(D-4)/2}$. This power behaviour is accompanied by
logs with the maximal power of $(n-1)$. The low-energy expansion of the partial wave amplitude $t_l^I(s)$ [details
of partial wave decomposition in $D$ dimensions can be found in Appendix~A] in LL approximation has the following form:

\begin{eqnarray}\label{A^I_D}
t_l^I(s)= \frac{2^{D-4}(4\pi)^{\frac D2}}{32\pi s^{\frac{D-4}{2}}}\sum_{n=1}^\infty \frac{\omega_{nl}^I}{2 l+D-3} {\hat S}^n
\ln^{n-1}\left(\frac{\mu^2}{s}\right).
\end{eqnarray}
We introduce here the dimensionless expansion parameter:

\be
\hat S =\frac{\sqrt \pi\ s^{k+\frac{D-4}{2}}}{(4\pi)^{\frac D2} 2^{D-4}\Gamma\left(\frac{D-3}{2}\right) F^2}.
\ee
Obviously, the expansion (\ref{A^I_D}) is reduced to Eq.~(\ref{t^I_lomega}) for $D=4$.

Now, with the help of Appendix~A the reader  can easily repeat the derivations presented in the Section~\ref{sec:general} and
after simple calculations arrive at the following recursion equation for the LL coefficients $\omega_{nl}^I$ for a $D$-dimensional theory:

\begin{widetext}
\begin{eqnarray}\label{w=ww_D}
\omega^I_{nl}=\frac{1}{n-1} \sum_{i=1}^{n-1}\sum_{l'=0}^{kn+(n-1)\frac{D-4}{2}}\frac 12\ \Big(\delta^{ll'}\delta^{IJ}+
C_{st}^{IJ} \Omega_{kn+(n-1)\frac{D-4}{2}}^{l'l}\left(D\right)
+C_{su}^{IJ}
(-1)^{l+l'}\Omega_{kn+(n-1)\frac{D-4}{2}}^{l'l}\left(D\right)\Big)\frac{\omega^{J}_{il'}\omega^{J}_{n-i,l'}}{2l'+D-3}.
\end{eqnarray}
\end{widetext}
The crossing matrices $\Omega^{ll'}_{n} \left(D\right)$ are straightforward
generalization of $\Omega_n^{ll'}$ for an arbitrary dimension:
\begin{eqnarray}
\nn
\Big(\frac{z-1}{2}\Big)^n C_l^{\frac{D-3}{2}}\Big(\frac{z+3}{z-1}\Big)=\sum_{l´=0}^{n}\Omega_{n}^{ll´}\left(D\right)C_{l´}^{\frac{D-3}{2}}(z)\, .
\end{eqnarray}
Here $C_l^{\frac{D-3}{2}}(z)$ are Gegenbauer polynomials, which form a basis for the partial wave decomposition of the amplitude in $D$ dimensions.
Details on the crossing matrices $\Omega^{ll'}_{n} \left(D\right)$ can be found in Appendix~B.

Various examples of multi-dimensional theories will be considered elsewhere. Here, as an example, we give only results for the LL coefficients
in the $6D$ $O(N)$ symmetric $\phi^4$ theory, which corresponds to $k=0, D=6$ in Eqs.~(\ref{A^I_D}, \ref{w=ww_D}).
The LL expansion for the partial wave amplitude in this theory has the following form:

\be
\label{expD6}
t_l^I(s)= \frac{2 \pi^2}{s}\sum_{n=1}^\infty \frac{\omega_{nl}^I}{2 l+3} \left(\frac{s}{128 \pi^3 F^2}\right)^n
\ln^{n-1}\left(\frac{\mu^2}{s}\right).
\ee
The four-loop results  for the coefficients $\omega_{nl}^I$ are presented in Tables~IV-VI of Appendix~C.

The recursion equation (\ref{w=ww_D}) for the $O(N)$ symmetric $\phi^4$ theory in $D$ dimensions can be solved in the large $N$ limit.
In this limit the amplitude is dominated by the S-wave and by the singlet $I=0$ ``isospin" component. Details of the large $N$ limit
for the recursion equations can be found in Ref.~\cite{MKV_LLog}. The result for the LL coefficients is:

\be
\omega_{n0}^{I=0}=-N \left(-\frac{N }{2 (D-3)}\right)^{n-1}\, .
\ee
With help of Eq.~(\ref{A^I_D}) we can perform the large-$N$ summation of LLs in the $D$-dimensional $O(N)$ symmetric $\phi^4$ model:
\be
t_0^0(s)&=&-\frac{N\sqrt\pi}{32\pi\Gamma\left(\frac{D-3}{2}\right)(D-3) F^2}\\
\nn
&\times&\frac{1}{1+\frac{N \sqrt \pi  s^{\frac{D-4}{2}}}{(4\pi)^{\frac{D}{2}}2^{D-3}(D-3)\Gamma\left(\frac{D-3}{2}\right)F^2}\ln\left(\frac{\mu^2}{s}\right)}.
\ee

\section{Conclusions}
Using the requirements of the unitarity, analyticity and crossing symmetry of the scattering amplitude, we derive
recursion equations for the coefficients in front of leading infrared logs in massless effective field theories.
The corresponding equations are given by (\ref{w=ww}) for $4D$ theories and by (\ref{w=ww_D}) for an arbitrary even dimension.
To implement these equations one needs to perform a calculation of tree diagrams only. One needs such calculation to find  initial conditions
for the recursion equations (\ref{w=ww},\ref{w=ww_D}).

In Section~\ref{sec:sigma} we demonstrate  that our recursion equations for LLs (\ref{w=ww}) are equivalent to the recursion equation
derived in Ref.~\cite{MKV_LLog} by a more complicated method. The method of Ref.~\cite{MKV_LLog} required non-trivial all-order analysis
of the structure of possible counter-terms as well as rather complicated loop calculations. In the present paper we found a much simpler
and more general way to derive the corresponding recursion equations.

In Section~\ref{sec:fi^4} we show that our recursion equations
(\ref{w=ww},\ref{w=ww_D})
are reduced to usual
RGEs for the case of a renormalizable field theory. Therefore, one can consider our recursion equation for LLs  as a generalization
of RGEs for the case of a non-renormalizable field theory.

Eq.~(\ref{w=ww}) in the case of a renormalizable field theory is reduced to a simple first order differential equation
(RGE). For a general case it can be reduced to a Hammerstein integral equation \cite{hammerstein}. For that type of integral equations
one can prove that the solution does exist and is unique. A little is known  about explicit solutions of the Hammerstein integral
equations, although many approximate methods have been developed. One of them is the reduction of the integral equation to the recursion equation
of the type (\ref{w=ww}). Indeed we checked on many examples of EFTs that the corresponding recursion equation is very effective and it allows one to obtain
LL coefficients to essentially unlimited loop order (e.g. it take a couple of minutes on a PC to compute the 99-loop LL coefficients for massless ChPT).

The derivation of our main equations is based on general properties of a quantum field theory: the unitarity, analyticity and the crossing symmetry.
The specific form of the theory enters the equation only through the crossing matrices (type of internal symmetry) and the
initial conditions for the recursion equation. That shows that the recursion equation for leading infrared logs has a general nature
and can be easily written for any massless EFT with fields of various spins (e.g. gravity or theory with spinor fields) and in an arbitrary dimension.

\section*{Acknowledgments}
We are thankful to N.~Kivel for many illuminating discussions. N.~Sverdlova is thanked for her help in preparation of the manuscript.
The work is supported in parts by German Ministry for Education and Research (grant 06BO9012) and by Russian Federal Programme ``Research and Teaching Experts in Innovative Russia"
(contract 02.740.11.5154).
\bigskip

\appendix
\section{Analytical properties of partial waves.}
In this appendix we give a summary of definitions and relations
for partial waves in arbitrary dimension $D$.

The partial waves in $D$ ($D-1$ spatial plus 1 time) dimensions are defined as
\begin{widetext}
\begin{eqnarray}\label{t_l^I}
A^I(s,t)&=&64\pi\sum_{l=0}^\infty
\frac{2l+D-3}{2}\frac{\Gamma\Big(\frac{D-3}{2}\Big)}{\sqrt{\pi}}
 C^{\frac{D-3}{2}}_l(\cos\theta)t^I_l(s) \\
\nn t_l^I(s)&=&\frac{1}{64\pi}\ \frac{\Gamma\Big(\frac{D-3}{2}\Big)}{\sqrt{\pi}}\frac{2^{D-4}l!}{\Gamma(l+D-3)}
\int_{0}^\pi d\theta\ \sin^{D-3}\theta
A^I(s,\cos\theta)C^{\frac{D-3}{2}}_l(\cos\theta),~~\cos\theta=1+\frac{2t}{s},
\end{eqnarray}
\end{widetext}
where $C^\nu_l(z)$ are Gegenbauer polynomials. Note that $C^{1/2}_l(z)=P_l(z)$, hence
the expansion (\ref{t_l^I}) for $D=4$ is reduced to the usual partial wave expansion in Legendre polynomials.
The unitarity of the $S$-matrix leads to the following relation for the
partial scattering amplitudes:
\begin{eqnarray}\label{Unitarity_D}
\Im
t^I_l(s)&=&\pi^{\frac{D-4}{2}}\left(\frac{s}{(4\pi)^2}\right)^{\frac{D-4}{2}}
|t_l^I(s)|^2\\
\nn
&+&\mathcal{O}(\text{Inelastic part})
\end{eqnarray}
This unitarity relation allows one to obtain the discontinuity of the amplitude
on the right cut ($s>0$) in the complex plane of the Mandelstam variable $s$.

The discontinuity on the left cut can be obtained with the help
of dispersion relations.
If one takes into account only cuts related to a
two-particles  intermediate state (that is enough for the LL approximation) the analytical properties of the amplitude are quite
simple. The amplitude has the $s$-channel cut from $4 m^2$ to
$+\infty$, and the $u$-channel cut from $0$ to $-\infty$. [We switch on
the masses for a moment in order to avoid the problems with
coalescing of branch points]. The usual dispersion relation at fixed $t$ with no
subtractions can be written  in the form, see e.g. \cite{Roy},
\begin{widetext}
\begin{eqnarray}\label{disp_Csu}
A^I(s,t)=\frac{1}{\pi}\int_{4m^2}^\infty ds'
\Bigg(\frac{\delta^{II'}}{s'-s-i0}+\frac{C_{su}^{II'}}{s'-4m^2+t+s-i0}\Bigg)\Im
A^{I'}(s',t),
\end{eqnarray}
\end{widetext}
where the matrix $C_{su}$ is the crossing matrix (\ref{Crossing}).
In principle, we have to make
subtractions in this dispersion relation, but the subtractions do not influence the imaginary
part of amplitude, and therefore for our consideration we can drop them. The discontinuity
of the amplitude (\ref{disp_Csu}) on the left cut ($s<0$) receives the contribution from the second term  of Eq.~(\ref{disp_Csu}) only.
Computing the corresponding discontinuity on the left cut and performing the partial wave decomposition
(\ref{t_l^I}) we find the following relation between discontinuities on the right and left cuts:
\begin{widetext}
\begin{eqnarray}\label{ImRoy_mass}
\Im
\,t^I_l(s)&=&\sum_{l'=0}^\infty C_{su}^{II'}\frac{2^{D-3}(2l'+D-3)}{\Gamma(l+D-3)}\frac{\Gamma^2\Big(\frac{D-3}{2}\Big)}{\pi}l!
\\ &\times& \int_{4m^2}^{4m^2-s} \frac{ds'}{s-4m^2}
\,\Big[\frac{4s'(4m^2-s-s')}{(s-4m^2)^2}\Big]^{\frac{D-4}{2}}
C^{\frac{D-3}{2}}_l\Big(\frac{s+2s'-4m^2}{4m^2-s}\Big)C^{\frac{D-3}{2}}_{l'}\Big(\frac{2s+s'-4m^2}{4m^2-s'}\Big)\Im\,
t_{l'}^{I'}(s').\nn
\end{eqnarray}
\end{widetext}
This relation for $D=4$ for the case of the $\pi\pi$ scattering
amplitude was derived in Ref.~\cite{Biss2} as a consequence of the Roy
equation. Taking $D=4$ and the limit $m^2\rightarrow 0$  we obtain
to the expression  (\ref{ImRoy}).

In the massless limit it is more convenient to rewrite Eq.~(\ref{ImRoy_mass}) in the form
\begin{widetext}
\begin{eqnarray}\label{ImRoy_D}
\Im
\,t^I_l(s)&=&-\sum_{l'=0}^\infty C_{su}^{II'}\frac{(2l'+D-3)}{\Gamma(l+D-3)}\frac{\Gamma^2\Big(\frac{D-3}{2}\Big)}{\pi}2^{D-4}l!
\\ &\times& \int_{-1}^{1} dz
\,(1-z^2)^{\frac{D-4}{2}}
(-1)^{l+l'}C^{\frac{D-3}{2}}_l(z)C^{\frac{D-3}{2}}_{l'}\Big(\frac{z+3}{z-1}\Big)\Im\,
t_{l'}^{I'}\big(\frac{s}{2}(z-1)\big).\nn
\end{eqnarray}
\end{widetext}

\section{Crossing matrices $\Omega_n^{ll'}$}
In $D$ dimensions the  definition of the crossing matrix in the
partial wave space is the following:
\begin{eqnarray}\label{Om_D}
\Big(\frac{z-1}{2}\Big)^nC_l^{\frac{D-3}{2}}\Big(\frac{z+3}{z-1}\Big)=\sum_{l'=0}^{n}\Omega_{n}^{ll'}(D)C_{l'}^{\frac{D-3}{2}}(z).
\end{eqnarray}
Using the orthogonality relations for the Gegenbauer polynomials, we can write the crossing matrix as the following integral:
\begin{widetext}
\begin{eqnarray}\label{Om_D(explicit)}
\Omega_{n}^{ll'}(D)=\frac{2l'+D-3}{2}\frac{2^{D-4}l'!}{\Gamma(l'+D-3)}\frac{\Gamma^2\Big(\frac{D-3}{2}\Big)}{\pi}
\int_{-1}^1 dz (1-z^2)^{\frac{D-4}{2}}
\Big(\frac{z-1}{2}\Big)^nC_l^{\frac{D-3}{2}}\Big(\frac{z+3}{z-1}\Big)C_{l'}^{\frac{D-3}{2}}(z).
\end{eqnarray}
\end{widetext}
This integral can be computed in terms of hypergeometric function in the Saalschutz form:
\begin{widetext}
\begin{eqnarray}\label{Om_F}
\Omega_{n}^{ll'}(D)=\frac{(-1)^{l'+n}(2l'+D-3)}{\Gamma(n+l'+D-2)}\frac{n!}{(n-l')!}\frac{\Gamma(l+D-3)}{A!}
~_4F_3 \left.\left(\begin{array}{c} -l,l+D-3,-l'-n-D-3,l'-n \\
-n,-n-\frac{D-4}{2},\frac{D-2}{2} \\\end{array}\right|1\right).
\end{eqnarray}
\end{widetext}
This representation of the crossing matrices $\Omega_{n}^{ll'}(D)$ is very convenient for the numerical calculations.

The main properties of the crossing matrices $\Omega_{n}^{ll'}(D)$ are the following:
\begin{eqnarray}\label{Om_prop}
&&\sum_{j=0}^n \Omega_{n}^{lj}(D)\Omega_{n}^{jl'}(D)=\delta^{ll'},\\
\nn
&&\sum_{j=0}^n (-1)^{j}\Omega_{n}^{lj}(D)\Omega_{n}^{jl'}(D)=(-1)^{l+l'}\Omega_{n}^{ll'}(D),\\
\nn
&&\Omega_n^{00}(D)=\frac{(-1)^n 2^{D-3}\Gamma\left(\frac{D-1}{2}\right)\Gamma\left(\frac{D}{2}+n-1\right)}{\sqrt \pi \Gamma\left(D+n-2\right)}.
\end{eqnarray}

\newpage

\section{Tables for LL coefficients}

In this Appendix we present Tables for LL coefficients for the $4D$ $O(N+1)/O(N)$ $\sigma$-model (Tables~I-III) and for the
$6D$ $O(N)$ symmetric $\phi^4$ model (Tables~IV-VI). The empty entries in the Tables correspond to zeros.

\begin{widetext}
\begin{center}
\begin{table}[h]\label{sigma_I=0(mod)}
\caption{Table of $I=0$ LL coefficients for the $4D$ $\sigma$-model,  $\omega_{nl}^{I=0}\cdot(N-1)^{-1}$}
\small{$
\begin{array}{c||l|l|l|l|}
n\setminus l & 0 & 2 & 4  \\ \hline
1 & 1 & ~ & ~  \\ \hline
2 & \frac{N}{2}-\frac{1}{9} & \frac{5}{18} & ~  \\ \hline
3 & \frac{N^2}{4}-\frac{61 N}{144}+\frac{59}{144} & -\frac{13N}{144}+\frac{13}{48} & ~  \\ \hline
4 & \frac{N^3}{8}-\frac{631 N^2}{2700}+\frac{46279 N}{194400}-\frac{13309}{194400} &
\frac{173 N^2}{2160}-\frac{4313 N}{38880}+\frac{5333}{38880} & \frac{N^2}{200}-\frac{49 N}{5400}+\frac{8}{675} \\ \hline
5 & \begin{array}{l}\frac{N^4}{16}-\frac{136 N^3}{675}+\frac{2498743 N^2}{7776000} \\ -\frac{3083771 N}{11664000}+\frac{619889}{4665600} \end{array}& \begin{array}{l}-\frac{1417 N^3}{40320}+\frac{481367N^2}{3628800} \\ -\frac{727373 N}{4082400}+\frac{1071107}{6531840}\end{array} & \begin{array}{l} -\frac{N^3}{280}+\frac{9787 N^2}{756000} \\ -\frac{449681 N}{27216000}+\frac{81007}{5443200}\end{array}
\end{array}$}
\end{table}

\begin{table}[h]\label{sigma_I=1}
\caption{Table of $I=1$ LL coefficients for the $4D$ $\sigma$-model, $\omega_{nl}^{I=1}$}
\small{$
\begin{array}{c||l|l|l|l|}
n\setminus l & 1 & 3 &5  \\ \hline
1& 1 & ~ & ~ \\ \hline
2& -\frac{N}{2}+\frac{3}{2} & ~ & ~ \\ \hline
3& \frac{9 N^2}{40}-\frac{37 N}{80}+\frac{49}{80} & \frac{N^2}{40}-\frac{37 N}{720}+\frac{49}{720} & ~ \\ \hline
4& -\frac{N^3}{10}+\frac{493 N^2}{1200}-\frac{41791 N}{64800}+\frac{8543}{12960} & -\frac{N^3}{40}+\frac{79 N^2}{900}-\frac{7019 N}{64800}+\frac{233}{2592} & ~ \\ \hline
5& \begin{array}{l}\frac{5 N^4}{112}-\frac{52859 N^3}{302400}+\frac{18963533 N^2}{54432000} \\-\frac{9585587 N}{27216000}+\frac{2685037}{10886400}\end{array} & \begin{array}{l}\frac{5 N^4}{288}-\frac{98743 N^3}{1555200}+\frac{4018577 N^2}{34992000} \\ -\frac{3612281 N}{34992000}+\frac{292127}{4665600} \end{array}& \begin{array}{l}\frac{N^4}{2016}-\frac{20753 N^3}{10886400}\\ +\frac{363091
   N^2}{97977600}-\frac{17849 N}{4898880}+\frac{101}{40824}\end{array}
\end{array}$}
\end{table}

\begin{table}[h]\label{sigma_I=2}
\caption{Table of $I=2$ LL coefficients for the $4D$ $\sigma$-model, $\omega_{nl}^{I=2}$}
\small{$
\begin{array}{c||l|l|l|l|}
n\setminus l & 0 & 2 & 4  \\ \hline
1& -1 & ~ & ~ \\ \hline
2& \frac{N}{3}+\frac{1}{9} & \frac{N}{6}-\frac{5}{18} & ~ \\ \hline
3& -\frac{N^2}{8}+\frac{3 N}{16}-\frac{59}{144} & -\frac{N^2}{8}+\frac{47 N}{144}-\frac{13}{48} & ~ \\ \hline
4& \frac{N^3}{20}-\frac{857 N^2}{10800}+\frac{21131 N}{194400}+\frac{13309}{194400} & \frac{N^3}{14}-\frac{449 N^2}{2160}+\frac{68711 N}{272160}-\frac{5333}{38880}
   & \frac{N^3}{280}-\frac{41 N^2}{3600}+\frac{407 N}{25200}-\frac{8}{675} \\ \hline
5& \begin{array}{l}-\frac{N^4}{48}+\frac{1727 N^3}{25920}-\frac{3323209 N^2}{23328000}\\+\frac{1492651 N}{11664000}-\frac{619889}{4665600} \end{array} & \begin{array}{l}-\frac{25 N^4}{672}+\frac{112891
   N^3}{725760}-\frac{4774289 N^2}{16329600}\\ +\frac{612299 N}{2041200}-\frac{1071107}{6531840} \end{array}& \begin{array}{l}-\frac{N^4}{224}+\frac{21797 N^3}{1209600}-\frac{1747919
   N^2}{54432000}\\+\frac{282487 N}{9072000}-\frac{81007}{5443200}\end{array}
\end{array}$}
\end{table}

\begin{table}[h]\label{D=6_I=0}
\caption{Table of $I=0$ LL coefficients for the $6D$ $\phi^4$-model, $\omega_{nl}^{I=0}$}
\small{$
\begin{array}{c||l|l|l|l|}
n\setminus l & 0 & 2 & 4  \\ \hline
1 & -(N+2) & ~ & ~ \\ \hline
2& \frac{1}{6}(N+2)(N-1) & ~ & ~ \\ \hline
3& -\frac{1}{36}(N+2)^2(N-1) & ~ & ~ \\ \hline
4& (N+2)^2(N-1)(\frac{N}{216}+\frac{1}{540}) & -\frac{7}{9720}(N+2)^2(N-1) & ~ \\ \hline
5& \begin{array}{l}-(N+2)^2(N-1) \\ \times(\frac{N^2}{1296}+\frac{223 N}{151200}+\frac{1}{226800})\end{array}
    & \begin{array}{l} -(N+2)^2(N-1) \\ \times(\frac{23 N}{388800}+\frac{101}{583200})\end{array} & \begin{array}{l}-(N+2)^2(N-1) \\ \times(\frac{N}{680400}+\frac{23}{5103000})\end{array}
\end{array}$}
\end{table}

\begin{table}[h]\label{D=6_I=1}
\caption{Table of $I=1$ LL coefficients for the $6D$ $\phi^4$-model, $\omega_{nl}^{I=1}$}
\small{$
\begin{array}{c||l|l|l|l|}
n\setminus l & 1 & 3 & 5  \\ \hline
1& 0 & ~ & ~ \\ \hline
2& \frac{1}{18}(N+2) & ~ & ~ \\ \hline
3& \frac{1}{108}(N+2)^2 & ~ & ~ \\ \hline
4& (N+2)^2(\frac{N}{756}-\frac{2}{2835}) & (N+2)^2(\frac{N}{15120}-\frac{1}{28350}) & ~ \\\hline
5& (N+2)^2(\frac{5 N^2}{27216}+\frac{N}{4320}-\frac{53}{136080}) & (N+2)^2(\frac{N^2}{45360}+\frac{N}{43200}-\frac{127}{2268000}) & ~
\end{array}$}
\end{table}
\clearpage{}
\begin{table}[h]\label{D=6_I=2}
\caption{Table of $I=2$ LL coefficients for the $6D$ $\phi^4$-model, $\omega_{nl}^{I=2}$}
\small{$
\begin{array}{c||l|l|l|l|}
n\setminus l & 1 & 3 & 5  \\ \hline
1& -2 & ~ & ~ \\ \hline
2& -\frac{1}{6}(N+2) & ~ & ~ \\ \hline
3& (N+2)(\frac{1}{18}-\frac{N}{60}) & -\frac{N}{540} & ~ \\ \hline
4& (N+2)(-\frac{N^2}{540}+\frac{N}{2700}-\frac{1}{270}) & (N+2)(-\frac{N^2}{2160}-\frac{29 N}{24300}+\frac{7}{4860}) & ~ \\ \hline
5& \begin{array}{l}(N+2)(-\frac{N^3}{4536}+\frac{19 N^2}{453600} \\  -\frac{17 N}{56700}+\frac{1}{113400})\end{array} & \begin{array}{l}(N+2)(-\frac{N^3}{11664}-\frac{77
   N^2}{388800} \\  +\frac{N}{97200}+\frac{101}{291600}) \end{array}& \begin{array}{l}(N+2)(-\frac{N^3}{408240}-\frac{N^2}{170100} \\ -\frac{N}{1701000}+\frac{23}{2551500})\end{array}
\end{array}$}
\end{table}
\end{center}
\end{widetext}

\end{document}